\documentstyle[twocolumn,prl,aps,epsfig]{revtex}

\begin{document}
\renewcommand{\thefootnote}{\fnsymbol{footnote}}
\sloppy

\newcommand \be{\begin{equation}}
\newcommand \bea{\begin{eqnarray} \nonumber }
\newcommand \ee{\end{equation}}
\newcommand \eea{\end{eqnarray}}
\newcommand{\rar}{\rightarrow}
\newcommand{\eq}{equation}
\newcommand{\eqs}{earthquakes}
\newcommand{\rp}{\right)}
\newcommand{\lp}{\left(}

\title{Analysis of the phenomenon of speculative trading\\
in one of its basic manifestations: postage stamp bubbles}

\author{B. Roehner$^1$ and D. Sornette$^{2,3}$} 
\address{$^1$  L.P.T.H.E., University Paris 7, 2 Place Jussieu \\
75251 Paris Cedex 05, France, roehner@lpthe.jussieu.fr\\
$^2$ IGPP and ESS Department\\
University of California, Los Angeles, California 90095\\
$^3$ LPMC, CNRS UMR6622 and Universit\'e des
Sciences\\
B.P. 70, Parc Valrose, 06108 Nice Cedex 2, France}
\date{\today}
\maketitle

\begin{abstract}

We document and analyze the empirical facts concerning one of the clearest
evidence of speculation in financial trading as observed in 
the postage collection stamp market. We unravel some of
the mechanisms of speculative behavior which 
emphasize the role of fancy and collective behavior. 
In our conclusion, we propose a classification of speculative markets based
on two parameters, namely the amplitude of the price peak and a second
parameter that measures its ``sharpness''.
This study is offered
to anchor modeling efforts to realistic market constraints and observations.

\end{abstract}
\vspace{5mm}


\narrowtext

\section{Introduction}

In his ``Treatise on general sociology'' \cite{Pareto},
Vilfredo Pareto pointed out that the construction of celestial
mechanics has been favoured by the fact that the mass of the sun is many
times larger than the masses of the largest planets. In other circumstances,
for instance with a double star in place of the sun or with a sun's mass 
only a few times larger than the mass of the largest planets, the movements
of the planets would be considerably more complicated. As a result, the 
three Kepler's laws would no longer hold; instead of a 2-body problem,
one would have to tackle a 3-or 4-body problem, which cannot be done
without a thorough understanding of non-integrable hamiltonian dynamics and
computer-assisted numerical computations. Under such 
conditions, the understanding of the laws of gravitation might have been 
delayed by at least two centuries.

In some respects, one is facing a similar difficulty in the analysis of 
financial markets as one has to deal with a many-body problem. First,
many investors are active in a typical trading day and their market impact
drives constantly the prices up and down. The difficulty is increased by the 
recent suggestion that the effective number $N$ of traders who
count on the market is not very large in the sense of the usual
``thermodynamical limit'' in physical systems (which usually provide
important simplification for modeling), probably
of the order of hundreds, as all models of market microstructure lead
to trivial deterministic dynamics when the limit of large $N$ is taken
\cite{StaufferlargeN}. 
Secondly, the many-body nature of the problem 
is further complicated by the interconnection between the
equity, bond, commodity and real estate
markets. This is shown by the following examples.
\begin{itemize}
\item In Vigreux et al. \cite{Vigreux}, one can find a spectacular example of the influence
of new bond emissions on the price level in the equity market: between 1954
and 1962, several large bonds have been issued at the Paris Stock Exchange which,
by absorbing a substantial part of available funds, brought down the equity 
market by as much as $20\%$ for the largest emissions. 
\item The connection
between the real estate market and the equity market has been illustrated 
in the early 1990s when the burst of the speculative bubble in Japan provoked
a parallel fall (of as much as 50 percent) in both markets  
\cite{StoneZiemba,Taniguchi,JohansorNikkei}. The recent
financial crises in Malaysia and Thailand also seem to have been triggered by
a fall in property prices \cite{eastern}. The role of intermediaries and of
herding has also been pointed out \cite{Krugman}.
\item It can be remembered that the Great Depression of
1929-1933 was, apart from the Stock market crash of Oct. 1929, marked  by a sharp decline in
wheat prices which in fact already started in 1925. 
\item These last years, one has witnessed that
the US stock exchange is very sensitive to rumors concerning interest
rates. Pushing the illustration, a sybillin remark from the president of the 
Federal Reserve suggesting a drop of the short-term rate is enough to trigger
important sell-out of bonds, with investment reported to stocks, leading to a surge of 
the Dow Jones lasting typically a full week. Inversely, when the Dow Jones drops, the
long-term interest rates fall down, which is a proof that the cash taken out
of the stock market has been carried over to the bond market. In a nutshell,
there is a kind of pendulum dynamics of the cash between the two markets.
\end{itemize}
To deal jointly with stocks, commodities and property is an awesome
perspective for this involves almost the whole economy either directly
or indirectly. 

Simpler phenomenologies appear when analyzing stock market price
fluctuations at short-time scales, from the tick scale (trade
to trade transaction time) to scales 
of about one month, for which the coupling between
different markets is less overwhelming, at least in normal circumstances,
and for which the structure may
be argued to be controlled in large part by simple market rules.
 Exponentially truncated L\'evy laws \cite{MantegnaStan,comment,bookBou}
with exponent around $\alpha \approx 1.5$ for the 6-year period 1984-1989
and power laws with exponents $\alpha \approx 3$ for the 2-year period 1994-1995 
\cite{exponentstanley},
superposition of Gaussian motivated by an analogy with turbulence
\cite{Gas,Arnemuzy} or stretched exponentials \cite{laher} have been proposed
to describe the empirical distribution of price returns in organized markets

Another strategy to simplify the problem is to study periods when financial
markets were still embryonic. This was the case before 1850;
since in addition wheat was before the 20th century by far
the most important commodity in Western Europe,
wheat price patterns can be expected to constitute
a fairly isolated phenomenon (with the
obvious qualification that they are influenced by meteorological factors). This
approach has been explored by Roehner \cite{Roehner1} and Roehner and Sornette 
\cite{RoehnerSor}.

In the present paper, we present an alternative empirical investigation which exemplifies
one single factor underlying market dynamics, namely ``speculation''.
In recent years, many groups have come up with interesting microscopic
models of stock market price dynamics that put emphasis on such an endogenous speculative
origin for the observed complexity 
of market prices 
\cite{Markovitz,Orlean,Taka,Levy,Bak,Calda,Johnson,BC,hierarmodel,Lux,Stau,Farmer,Ilinski,Iori}.
Here, we present what we consider to be probably the purest case illustrating 
speculation in a market, as it occurs in the collector's stamp market, just like
the motion of planets was for Kepler and Newton the purest case of frictionless motion.
This market has a number of definite advantages in 
terms of simplicity. 
\begin{enumerate}
\item It is relatively isolated from other speculative markets
because the proportion of the collectors is by far larger than that of 
the investors.
\item ``Production'' and ``consumption'' take on particularly simple forms: production
is restricted to  a short time span and the production figures are statistically
well known; since most collectors' stamps are not actually used on letters,
consumption is basically non-existent; it only occurs by wear and tear or
by accident at a small and probably fairly constant rate. 
\item In contrast to gold, silver or copper coins, stamps cannot be melted.
A few decades after they have been issued, 
they can no longer be used and have therefore no intrinsic value; 
in other words, their prices are
solely determined by the judgment of the collectors.
\item In contrast to other
collectibles such as paintings or furniture, stamps are fairly liquid assets. 
Any valuable stamp can be sold to a trader at a price given in the current 
catalogue (a discount might be applied which takes into account the state of 
conservation of the stamp).
\item Stamp markets display huge price bubbles. 
Multiplication of the current price by a factor of about $10$ within a decade is
not uncommon.
\item Stamp prices range from a fraction of a dollar to several 
thousand dollars. This gives the opportunity to observe the speculative behavior
of collectors when they are confronted with stakes of different magnitudes. 
\item The identification of what is a speculative bubble in the stamp market does not suffer from 
the same uncertainties as in other markets. Indeed, in recent years, an active debate
between economists has been aimed at the problem of an unambiguous and rigorous 
definition of speculative bubbles, by trying to distinguish those price increases due to 
changes in fundamentals from those resulting from pure speculation \cite{debatespecu}. 
The challenge stems for the fact that this
question is rather ill-posed in general because one does not know 
and does not have access to all relevant fundamentals. For instance, should the 
construction of the Opera-Bastille theatre be incorporated in the list of 
fundamentals defining the real-estate market in the 11th Paris district?
In the stamp market, there are very few fundamentals and they are well-known.
The definition of a speculative bubble is thus much clearer.
\end{enumerate}

To be fair, one has to recognize that, as far as its statistical analysis is 
concerned, the stamp market also has a number of drawbacks. First, stamp catalogues
are published only every year (sometimes even every two or three years). As the 
catalogues are the only practical mean for knowing the prices of stamps in a fairly
systematic way, this precludes any investigation of short term fluctuations.
The bulk of stamp transactions takes place between private individuals; as 
a result it is almost impossible to estimate the volume of the transactions.

Let us now explain how an exploration of the stamp market may provide clues for
a better understanding of the mechanisms and patterns of speculation. Generally
speaking, it may be argued that several kinds of agents participate in a given
market. For instance, the operators in real estate markets can be divided into 
two subgroups. (i) Residents who buy and sell for their personal use and (ii) 
speculators or property developers who make money by selling and buying property. 
As an illustration, the later group represented about
$20\%$ of the buyers in the 1997 Paris real estate market (La Vie Fran\c caise 1998,
No 27589,9). The collective behavior of the residents obviously will not be the
same as that of the speculators. Yet, only the combined result of their actions
is accessible to observation. No doubt that such an intermingling of different
mechanisms markedly contributes to blur and obscure the interpretation of the 
phenomenon. Roehner \cite{Roehner1999} has tried to separate 
out residents and speculators by investigating the price bubble at the level
of separate districts. The proportion of the speculators turned out to vary
within a 1:2 margin. In the present paper, we pursue the same objective but,
instead of looking at different districts, we are going to consider different 
stamps. In collector's circles, some stamps are known to be speculative assets; 
examples will be given below. In summary, comparing the price evolution 
of different stamps may give us an insight into the collective behavior of
different populations of economic agents. 

Throughout this paper, we consider mainly, though not
exclusively, the market of non-used French stamps. The restriction to new stamps
is made because their quality and therefore their price are much easier to control. 
At a time of 
rapid worldwide internationalization, it could seem more surprising to restrict
ourselves 
to French or British stamps. At the collectors's level, there is a strong
force that works against internationalization. In the past half-century, the number
of sovereign countries has been multiplied by a factor three. Furthermore in most
countries, new stamps have been issued at a much faster rate than before World
War II; as a result there has been a huge increase in the number of stamps; this 
is reflected in the growing size of worldwide stamp catalogues; for instance the 
French Yvert and Tellier catalogue, which used to be in two volumes, now
has no less than eight volumes. In the face of such a bewildering diversity,
it is not surprising that collectors
are more and more tempted to restrict themselves to only one country or group
of countries. This is also in agreement with the typical collector's psychology of
specialization to a narrow niche that suits his/her fancy.

The paper is organized as follows. In the second section, we provide
estimates for the size of the stamp market and we discuss the question of the
reliability of the prices given in the catalogues; we also sketch the long-term
evolution of the stamp market. Section 3 provides some selected examples 
of price bubbles which shed light on the nature of speculative forces. In section 4, 
we propose a tentative classification of speculative markets according to 
the value taken by two important parameters, namely the amplitude of the price
peak and a second parameter which summarizes the form of the peak.

\section{The stamp market}

\subsection{Turn over}

Compared to stock or real estate markets, the markets of stamps for
collection are small. 
The French market can serve to illustrate this point. 
Approximately 40 stamps for collection are issued every year.
Sales of these newly issued stamps for collection can be estimated (for 
1984) to be of the order of 400 million francs (less than 
US\$80 millions) \cite{Massacrier}. Of this, about no more thant $5\%$
are used for mailing. This results from two factors: (i) the issued
stamps for collection have facial values that very rarely correspond to 
mailing values; (ii) the state of conservation of a stamp for collection
is so determinant for its value (discount for less than perfect conservation
can reach $50\%$ or more), that few collectors take the risk to use
these stamps for mailing. The 400 million francs issue of
stamps for collection for 1984 has only very slowly varied over the years,
being 420 million francs for 1993. This must be compared to the total value
of 5.5 billions francs in 1993 of stamps issued for mailing.

As we already noted, it is
more difficult to estimate the other transactions. The turn over of the 
five main traders was of the order of 300 million francs. If the total 
transaction figure is assumed (somewhat arbitrarily) to be four times larger,
one obtains an overall figure of less than 2 billion francs. This is larger
than 400 million francs because it comprises trading of all previously issued stamps. 
Let us compare 
this figure to the transactions on stocks, on real estates or on works of art.
\begin{itemize} 
\item By 1984, the annual transactions on the Paris Stock Exchange were of the order 
of 100 billions francs.
\item In 1984, 35000 appartments have been sold in Paris 
(figure given by the Chambre des Notaires, i.e. the Lawyers Association);
at an average price of one million francs per appartment, this represented an amount of 
35 billions francs.
\item The turnover of public auctions in works of art was in
1975 of the order ot one billion francs, while private transactions were
estimated at about 1.5 billion francs \cite{Peyrelevade}. 
\end{itemize}

The French stamp market thus represented in the 1980s about 2\% of the transactions on the 
stock market, about 6\% of the real estate sales in the city of Paris,
(i.e. excluding the suburbs); they were approximately of the same magnitude as 
those of works of art. 

\subsection{Estimating the price of stamps}

The stamp catalogues provide the prices of all existing stamps. In countries
such as Britain, France, Germany and the United States, such catalogues have
been published annually for more than a century. They thus constitute a 
valuable source of information for anyone who wants to study either price 
bubbles or the long term trend of stamp prices. However, the question arises as to
whether the prices given in the catalogues truly reflect the 
prices in actual transactions. From a collector's perspective, this is 
a complex question; yet from a statistical point of view, it will be seen
to have a simple answer. 

The prices listed in a catalogue are for stamps in a perfect state of 
conservation. It is however obvious that a stamp that has been bought several 
decades ago can hardly be in a perfect state. In other words, its price will
always be less than listed in the catalogue. Statistically however, there is 
a close connection between negociated prices and the prices listed in the
catalogue. This has been proved by Feuilloley \cite{Feuilloley} using a sample of 300 stamps; 
the correlation was about 0.90, while the regression coefficient was
about 0.5 which means that on average the real prices were only half the prices
listed in the Yvert and Tellier catalogue. In the following,
we are interested  in the evolution of relative prices rather than in their
absolute magnitude. The catalogues can thus be considered as a reliable source.

\subsection{Long-term trend of French stamp prices}

Fig.1 shows the long term price trend of (i) Nineteenth century stamps (ii) All
stamps listed in the Yvert \& Tellier catalogue. The deflated
price of 19th century 
stamps has increased at an average rate $r_{19}=5.2\%$, while the average
price of all stamps has grown at a rate $r_{\rm all} = 2.1\%$. 

The rate for 19th century stamps is easier to rationalize
than the second one for all stamps, since it concerns the evolution of
a sample of stamps which remained unchanged in the course of time. During the 
same time span, the net national income (at constant prices) has increased at an
annual rate of $r_0=3.1\%$. The difference $r_{19} - r_0 = 2.1\%$
can be interpreted with the following simple model.
\begin{itemize}

\item In the course of time, the offer, i.e. the number of
19th century stamps, has decreased at a constant rate of $d \%$. 
Thus, the residual number of stamps after a time $t$ is
$N_0 e^{-d t}$, starting from an initial number $N_0$.
One can advance the following rational for this decrease.
There are two types collectors\,:
\begin{enumerate}
\item the 'amateurs', those who have just a small collection. At some
point in time they stop collecting, and as their collection's worth is
low, the stamps will be thrown away.
\item The `professionals': they collect `seriously', and their collection
has a certain value, and even when they die their relatives will be
aware of the collection's value and sell it. Thus, the stamps 
re-enter the market.
\end{enumerate}
What makes things difficult is that the amateurs will normally not
have the rare stamps. Thus, the rare stamps'depreciation rate
is much flatter than that of the everyday stamps. For the rare stamps,
 the depreciation rate is probably rather close to zero.

\item On the demand side, one must consider the total amount of money 
$M$ that the total 
number of collectors $C$ are willing to devote to purchasing 19th century 
stamps. Let us denote by $c$ the proportion of collectors in the total
population $N$, and by $f$ the fraction of his/her revenue $R$
that a collector
is willing to spend on 19th century stamps. One has:
\be
 C=cN~, \qquad M = f R C = (NR) c f = f c I~, \label{jhakjak}
 \ee
where $ I $ denotes the national income.  
\end{itemize}

Within this simple framework, the difference $r_{19} - r_0 = 2.1\%$
can be attributed to the following factors. 
\begin{itemize}
\item The proportion $c$ of collectors in the total population has increased.
\item The number of 19th stamps has slightly decreased.
\item The proportion $f$ of a typical collector's revenue spent 
on 19th century stamps has increased. This conjecture seems quite 
reasonable in the light of Engel's law which states that, as
per capita income increases, the percentage spent on items other than
food, clothing or housing increases too. 
\item The number of stamps decrease because of their finite ``half-time''.
\end{itemize}

Equating (\ref{jhakjak}) to $N_0 e^{-d t} p(t)$ and using $I = I_0 e^{r_0 t}$
gives the estimation
\be
r_{19} = r_0 + d~.  \label{jfjal}
\ee
Actually, the equality should be replaced by the inequality $\geq$, to
take into account that the fraction $f$ of the revenue and the
proportion $c$ of collectors may have also increased. This gives a lower bound for
the half-time ($\ln 2 /d$) of about thirty years. If in addition, we incorporate a risk factor,
requiring that the rate $r_{19}$ of stamp price appreciation should include the price
of risk of the stamp destruction, typically proportional to the standard deviation
of the Poisson process of stamp destruction, this doubles the lower bound of the expected
half-time for 19th century stamps to sixty years. Indeed, including the price of risk
means that there must be a remuneration resulting
from the fact that the process is not certain 
and exhibits fluctuations. In this framework, the interest rate $r_{19}$
must incorporate a remuneration which is typically proportional to the risk, measured
usually by the standard deviation of the uncertain process. For a Poisson process,
the standard deviation is $1/d$. Adding this contribution, $d$ is replaced by $2d$ in
(\ref{jfjal}) and the corresponding estimation for $d$ is halved, hence the doubling
of the half-time.

If the interpretation provided
by this model is correct, one would expect the price of 19th century stamps
to continue to increase at a faster rate than the net national income. Since 
in addition, there are no taxes on stamp sales, stamps are likely to constitute a good 
investment for the foreseeable future. Notice that the 
difference $r_{19} - r_0 = 2.1\%$ between the rate of return of 19th stamps
and the net national income does not reflect the influence of a risk factor but rather 
that of a shift on the supply-demand curve towards increasing scarsity of supply
and increasing demand for old and rare stamps.

\section{Speculative bubbles}

In this section, we address the following questions\,:
\begin{enumerate}
\item Has the price of an item a determining influence on the way a speculative bubble
unfolds? 
\item Are there different speculative patterns? 
\item Is it possible 
to predict at least the upper bounds for the amplitude of the price peaks? 
\item {\it Why} does a specific stamp become the target of a speculative process?
\end{enumerate}

\subsection{How versus why}

The first three questions have to do with {\it how} the speculative process
develops. In other words, given that a speculative process has taken place, one
tries to analyse its characteristics; in contrast, the fourth question 
refers to the {\it why}'s. In a previous paper \cite{RoehnerSor},
we have already emphasized that this latter question is very difficult to address
specifically and this difficulty is a clue for the origin of bubbles.
As an illustration, consider the two
stamps that have been issued the same day (27 Oct. 1979) with the same number of
copies (6 millions) and with the same face value (2 francs). They only differed
by their themes: one (No 2059 in the C\'er\`es catalogue) represented an
ancient painting of ``Diane at the bath'', while the other (No 2060) represented
a painting by Van Gogh. As can be seen in Fig.2, only the second stamp
experienced a speculative process which multiplied its price by 10 in less than 
six years. 

When invited to 
comment on that enigma, a stamp expert explained that the speculation probably
started when a big collector (or trader) happened by chance to come in 
possession of a substantial proportion of the total number of stamps that had
been issued. In fact, many different rumours circulated in philatelic circles as
to how this happened. One possible explanation is that speculation seemed, at least 
in that case, to be triggered by purely subjective factors. Another scenario
explored by Roehner and Sornette \cite{RoehnerSor} (see also \cite{multipli})
is that chance plays a crucial role is nucleating the bubble which is then
amplified by multiplicative effects \cite{multipli} and/or path dependent positive
feedback effects \cite{Arthur}. In this last scenario, the initial price
inflation of the ``Van Gogh'' stamp was a ``lucky'' event, or maybe even 
the act of speculation by a big collector, which was afterwards amplified
by the action of positive feedbacks as described for instance in the Polya Urn model
(see \cite{Marsili} for a modern extension to describe self-organization): 
the bubble fed on itself, reinforcing itself by the increasing attraction
presented by the ``Van Gogh'' stamp. A similar scenario has been documented
for the real estate bubble that culminated in 1991 in France \cite{Friggit}\,:
a booming real-estate market is attracting to everybody but the poor who
cannot enter the market: sellers
cash in a substantial profit; buyers are not frightened by the 
astronomical price and buy confidently with the expectation 
that they will also cash a profit when they sell in the future.  This is 
sustainable only as long as there is liquidity, i.e. a reservoir of 
potential buyers is continuously replenishing.

\subsection{Impact of initial price levels}

We now turn to the first question, namely what is the influence of the price
level. To a large extent, we find that the price is irrelevant. 
In other words, an expensive stamp and a cheap stamp, which both became the targets 
of a speculative process, experience parallel price trajectories. However, there is
a low but statistically significant correlation between the price level and the 
price amplification factor. This fact is illustrated by Fig.3 for 
French stamps and by Fig.4 for British stamps. Fig.3 shows the evolutions
of a very rare stamp and of a fairly common stamp. In 1904, the price of stamp No 2
was 200 francs while the price of stamp No 16 was 0.05 francs. In spite of such 
a large price gap, the price evolutions are fairly similar. It is true 
that the timing was not the same, with the bubble for the cheap 
stamp beginning to build up about 10 
years earlier; but the overall increase was of the same order
as well as the subsequent decrease. Fig.4 provides a similar example. In 1965,
stamp No 90 was worth 4000 francs against 60 francs for stamp No 155. Yet, the price 
evolutions are very similar. In fact, almost all British stamps issued
before World War II and having a face value of a pound or more followed a 
similar evolution. Stamp No 106 provides an example of a stamp issued in 1902-1910
but having a face value of only half a penny; in this case, the speculative 
increase is very small in comparison. 

We now examine if there is a correlation between the
initial price of a stamp and its price amplification factor. The results are
presented in table 1a. In this table and the others below, 
the coefficient of amplification are given in current
value, following the habit of professionals.
The coefficient of linear correlation between the logarithms of 
the 1965 prices and the price amplification factor is equal to 0.55; in other
words, the higher the initial price the stronger the speculation seems to be.
A similar observation was made for the district-level prices during the
Paris real estate bubble \cite{Roehner1999}.
When the bubble started in 1984, the
price in the most expensive district (6th) was twice as high as the price in 
the cheapest one (10th). The bubble first began to built up in the most 
expensive districts (6th, 7th, 16th, 17th) and then spread to cheaper districts
with a delay of about 6 to 12 months. Furthermore there was a low (but nevertheless
significant) correlation of 0.49 between the 1984 prices and the price 
amplification factors as shown in table 1b.

\subsection{Speculative patterns}

\subsubsection{Corners}

We now turn to the second question, namely whether different speculation 
patterns can be identified. Table 2 shows that there is a marked difference
in terms of price increase rate between the first and the second half of the 
table. Let us for instance consider more closely one of the episodes 
in the second group, namely the
bubble which concerned a few French stamps in the mid 1980s. Owing to its
short duration and to its high increase rate, one may wonder whether this
was not a deliberate attempt to corner the market, i.e. to take the control
of the market by buying all available stamps. This question clearly raises
two other ones: what is the percentage of the total ``production'' of a stamp 
that it is necessary to buy in order to control the market 
and create such a ``squeeze''. Is it within the financial capability 
of big operators? Let us consider
the Van Gogh stamp (C\'er\`es No 2060); its face value is 2 francs and
6 millions copies have been issued representing a total amount of 12 million 
francs, that is to say about 3 percent of the total annual value of newly issued
stamps or 4 percent of the annual turnover of the five major
French traders. There should therefore be no problem for a trader or an 
important collector to buy at least 75 percent of the 6 million stamps. Once 
the bubble has reached its peak-level however, it becomes much more
difficult to keep the market under control for the 6 million stamps now 
represent $ 60\times 6= 360 $ millions francs, an amount which is 
of the same
magnitude as the global turnover of the five major traders. This has two consequences:
\begin{enumerate}
\item There is obviously an upper bound to the price level that can be
reached during a speculative bubble; this ceiling price is determined both
by the ability of the main operators to control the market and by the 
subsequent ability of the market to absorb the offer. It seems that 60 francs
for the Van Gogh stamp was either close to or even beyond this upper bound.
\item The 360 millions francs represent about 20 percent of our estimate
for annual transactions; clearly the market is not going to devote 20 
percent of its purchasing power to just one stamp among several thousand other
French stamps. In other words, the market will obviously be 
unwilling to absorb the 6 million stamps (or even a substantial fraction of
them) at such a high price. The price is therefore bound to decrease before 
large stocks of stamps can be absorbed by the market. 
\end{enumerate}

\subsubsection{Shape of the price peak}

In order to characterize the shape of the bubble peaks, we use the quantification
defined by Roehner and Sornette \cite{RoehnerSor} describing the price $p(t)$
as a function of time according to\,:
\be
p(t) = a \exp \left[ -\hbox{sgn}(\tau )\left| {  t-t_0
\over \tau }\right| ^{\alpha} \right]  ~,
\label{eqe}
\ee
where $ t_0 $ denotes the turning point of the peak and $\tau$ is a characteristic
time scale for the maturation of the bubble.
 The key parameter that 
quantifies the shape of the peak is the exponent $\alpha$.
\begin{itemize}
\item If $ \alpha $
is equal to $ 1 $, one retrieves an exponential growth up to the turning
point followed by an exponential decay. $x = \ln(p)$ is thus linear up to the maximum,
with a tent-like structure.
\item If
$ \alpha <1 $ and $ \tau > 0 $, the function describes a
sharp peak (accelerating rise before the peak and decelerating drop after the peak). 
\item If $ \alpha >1 $ and $ \tau <0 $, the function
describes a flat trough (decelerating drop followed by an accelerating rise).
\item If $ \alpha >1 $ and $ \tau >0 $, the
function describes a ``flat peak'' (decelerating rise followed by an accelerating drop).
\item If $ \alpha <1 $ and $ \tau <0 $, the function describes a
sharp trough (accelerating drop followed by a decelerating rise).
\end{itemize}

Table 2 shows that almost all price peaks that we examined have a peak exponent
$\alpha > 1$. This corresponds to a flat peak pattern. This must be
contrasted to the situations found for most commodity price
peaks for which a sharp peak-flat through pattern holds \cite{RoehnerSor}. The only stamp
bubble which clearly displays a sharp peak pattern is the one that occurred
in France during World War II. The ``causes'' of this peak are relatively easy 
to enumerate, namely 
\begin{enumerate}
\item the need to sweep black market profits under the carpet,
\item the demand generated by a number of passionate collectors belonging to the
German occupation troops,
\item the fact that during the war, savings accumulated
as a result of consumption restrictions. 
\end{enumerate}
Yet, none of these reasons explains why the shape of these peaks was so much different.

\section{Conclusion}

By analysing the stamp market, we have tried to document a vivid
demonstration of speculation and of its main characteristics in a one 
of its most basic manifestation. Our observations
emphasize the role of fancy and collective behavior. With J.M. Keynes and 
his parallel between the stock market and newspapers'
beauty contests (see section 12 of the General Theory \cite{Keynes}),
one could argue that the main charateristic of the 
successful speculator is his/her ability to predict what the (average) behavior 
of the rest of the public will be. However, such a model probably underestimates 
the importance of social interaction and mimetic contagion. For instance, one
can hardly expect an isolated collector to be capable of predicting that,
among hundreds of other stamps, the Van Gogh stamp will become the target of an
important speculation movement. Such a behavior is more likely to be propagated
by the interactions between collectors and by philatelic publications. 
Unfortunately, lack of data prevents us from statistically estimating the 
strength and frequency of those social interactions. 

In order to provide a unified overview of some of the results presented 
in this paper and in \cite{RoehnerSor} and to compare the shape of bubbles
found in commodities and in collecting items, Fig.5 shows
comparison between different types of speculative movements. 

A question left open by our present study is the possible correlation 
between the structure and shape of the speculative bubble and the 
degree of market liquidity. By this, we mean the following. 
Speculators are very useful to provide ``liquidity'' in the market. 
Otherwise, in the stamp or real-estate markets, exchanges would occur 
only between collectors or residents and would be very scarse since
the purpose of a collector is to collect and that of a resident is to reside!
The task of collecting and changing living place would thus be proportionally
more difficult. This positive role of speculation is classical and is often
advanced to defenders of free capitalistic markets. On the other hand,
a larger proportion of speculators imply that traders are more in phase, there is
less friction and the speculative bubbles should develop faster. This could be 
tested empirically by correlating the exponent $\alpha$ to the ratio of 
residents to speculators in the 20 Parisian districts. 

Our results on the independence of the shape of the speculative peaks 
with respect to the price of the stamps
suggest that risk aversion (related to the amount of money
involved in a transaction) does not play an important role in the speculative 
bubbles observed in the stamp market. This is a very interesting information,
whose validity should be investigated for other markets, including financial
markets. Theoretical models of financial crashes using rational expectation theory
coupled to herding behavior of a fraction of the traders suggest also that risk 
aversion is not a determining factor \cite{crash}. It is thus reassuring that
a similar conclusion is obtained on two very different markets and by very different
approaches.

\vskip 1cm
Acknowledgements: We would like to express our gratitude to 
Mrs. Paganini and Luppi  (C\'eres-Philat\'elie) and to the librarians 
of the Documentation Center
for Collectors Stamps (Mus\'ee de la Poste-France Telecom).

\end{document}